\documentclass[reqno]{amsart}

\usepackage{amsmath}
\usepackage{amsfonts}
\usepackage{amsthm}

\numberwithin{equation}{section}
\newcommand {\pa}{\partial}

\newcommand{\meas}{\operatorname{meas}}
\newcommand{\supp}{\operatorname{supp}}
\newcommand{\Spec}{\operatorname{Spec}}

\newcommand{\Div}{{\operatorname{div}\,}}

\newcommand{\dist}{{\operatorname{dist}}}

\newcommand{\curl}{{\operatorname{curl}\,}}

\newcommand{\Hess}{{\operatorname{Hess}\,}}

\newtheorem{theorem}{Theorem}[section]
\newtheorem{thm}[theorem]{Theorem}
\newtheorem{lemma}[theorem]{Lemma}
\newtheorem{proposition}[theorem]{Proposition}
\newtheorem{prop}[theorem]{Proposition}

\newtheorem{remark}[theorem]{Remark}

\newtheorem{cor}[theorem]{Corollary}

\newtheorem{assumption}[theorem]{Assumption}

\title{On the Ginzburg-Landau critical field in three dimensions}

\author{S. Fournais}
\author{B. Helffer}

\address[S. Fournais]{Department of Mathematical Sciences, University of Aarhus, Ny Munke\-gade,
Building 1530,
DK-8000 Aarhus C, Denmark
}
\email{fournais@imf.au.dk}

\address[B. Helffer]{Laboratoire de
Math\'{e}matiques UMR CNRS 8628\\ Universit\'{e} Paris-Sud - B\^{a}t 425\\
F-91405 Orsay Cedex\\ France.
}
\email{bernard.helffer@math.u-psud.fr}
\date{\today}

\begin{document}

\bibliographystyle{plain}

\begin{abstract}
We study the three dimensional Ginzburg-Landau model of
superconductivity. Several `natural' definitions of the (third)
critical field, $H_{C_3}$, governing the transition from the
superconducting state to the normal state, are considered. We analyze
the relation between these fields and give conditions as to when they
coincide. An interesting part of the analysis is the study of the
monotonicity of the ground state energy of the Laplacian, with
constant magnetic field and with Neumann (magnetic) boundary
condition, in a domain $\Omega$. It is proved that the ground state
energy is a strictly increasing function of the field strength for
sufficiently large fields. As a consequence of our analysis we give an
affirmative answer to a conjecture by Pan.
\end{abstract} 

\maketitle
%% \tableofcontents

\section{Introduction}
{\bf In the whole paper  $\Omega \subset {\mathbb R}^3$ will
 be a bounded simply connected domain whose boundary is connected and Lipschitz continuous.}\\
Let $\beta$ be the constant magnetic field along the $z$ axis: $\beta = (0,0,1)$.
The Ginzburg-Landau functional in three space dimensions is given by
\begin{multline}
\label{eq:GL_F}
{\mathcal E}[\psi,{\bf A}] = {\mathcal
E}_{\kappa,H}[\psi,{\bf A}]  =
\int_{\Omega} \Big\{ |p_{\kappa H {\bf A}}\psi|^2 
- \kappa^2|\psi|^2
+\frac{\kappa^2}{2}|\psi|^4\Big\} \,dx\\
+ \kappa^2 H^2 \int_{{\mathbb R}^3}
|\curl {\bf A} - \beta|^2\,dx\;,
\end{multline}
with
$\psi \in W^{1,2}(\Omega;{\mathbb C})$, ${\bf A}$ in the space $\dot{H}^1_{{\bf F}, \Div}$ that we will define below, and where $p_{{\bf A}} = (-i\nabla+ {\bf A})$. 
Notice that the second integral in \eqref{eq:GL_F} is over the entire space, ${\mathbb R}^3$, whereas the first integral is only over the domain $\Omega$.

Let ${\bf F}$ be the vector potential 
\begin{align}
{\bf F}(x_1,x_2,x_3) = \tfrac{1}{2}(-x_2, x_1,0)\;.
\end{align}

Formally the functional is gauge invariant. In order to fix the gauge, we will impose that vector fields ${\bf A}$ have vanishing divergence. Therefore, a good choice for the variational space for ${\bf A}$ is 
\begin{align}
\dot{H}^1_{{\bf F}, \Div} = {\bf F} + \dot{H}^1_{\Div}\;, \end{align}
where
\begin{align*}
\dot{H}^1_{\Div} = \{ {\bf A} \in \dot{H}^1({\mathbb R}^3, {\mathbb R}^3) \,\big |\, \Div {\bf A} = 0 \}\;.
\end{align*}
We use the notation $\dot{H}^1({\mathbb R}^3)$ for the homogeneous Sobolev spaces, i.e. the closure of $C_0^{\infty}({\mathbb R}^3)$ under the norm
$$
f \mapsto \| f \|_{\dot{H}^1} = \| \nabla f \|_{L^2}\;.
$$
We will recall below the fact that any square integrable magnetic field ${\bf B}$, i.e. any vector field ${\bf B} \in L^2({\mathbb R}^3; {\mathbb R}^3)$ with $\Div {\bf B} = 0$ in the sense of distributions, can be represented by a vector field ${\bf A} \in \dot{H}^1_{\Div}$.

Minimizers, $(\psi, {\bf A}) \in W^{1,2}(\Omega) \times \dot{H}^1_{{\bf F}, \Div}$, of the functional ${\mathcal E}$ have to satisfy the Euler-Lagrange equations:
\begin{subequations}
\label{eq:GL}
\begin{align}
\label{eq:equationA}
p_{\kappa H {\bf A}}^2\psi =
\kappa^2(1-|\psi|^2)\psi & \quad \text{ in } \quad\Omega\;,\\
\label{eq:equationB}
\curl^2 {\bf A} =\big\{-\tfrac{i}{2\kappa H}(\overline{\psi} \nabla
\psi - \psi \nabla \overline{\psi}) + |\psi|^2 {\bf A}\big\} 1_{\Omega}(x)
&\quad \text{ in } \quad {\mathbb R}^3 \;,\\
(p_{\kappa H {\bf A}} \psi) \cdot N = 0 &\quad \text{ on } \quad \partial\Omega\;,
\end{align}
\end{subequations}
where $N(x)$ is the unit interior normal at the boundary.

It is not completely standard (but see \cite{Gi}
 for the analysis of this case) but rather easy 
 to prove that, for all $\kappa, H>0$, the functional ${\mathcal E}_{\kappa,H}$ has a minimizer. It is a result of Giorgi and Phillips, \cite{Giorgi-Phillips}, that for $\kappa$ fixed and $H$ sufficiently large (depending on $\kappa$), the unique solution of \eqref{eq:GL} (up to change of gauge) is the pair $(\psi, {\bf A}) = (0, {\bf F})$. Since $\psi$ is a measure of the superconducting properties of the state of the material and ${\bf A}$ is the corresponding configuration of the magnetic vector potential, the result of Giorgi and Phillips reflects the experimental fact that superconductivity is destroyed in a strong external magnetic field. 

We define the lower critical field, $\underline{H}_{C_3}$ as the value of $H$ where this transition takes place:
\begin{align}
\underline{H}_{C_3}(\kappa) := \inf\{ H>0 \;:\; (0, {\bf F}) \text{ is a minimizer of } {\mathcal E}_{\kappa,H}\}\;.
\end{align}
However, it is far from obvious from the functional that the transition takes place at a unique value of $H$---there could be an interval of transitions back and forth before the material settles definitely for the normal state, $(0,{\bf F})$. Therefore, we introduce a corresponding upper critical field
\begin{multline}
\overline{H}_{C_3}(\kappa) :=  \inf\{ H>0 \;:\; \text{for all } H'>H, \\
(0, {\bf F}) \text{ is the unique minimizer of } {\mathcal E}_{\kappa,H'}\} \;.
\end{multline}
Notice that our space $\dot{H}^1_{{\bf F}, \Div}$ fixes the choice of gauge.

It is the objective of this paper to give conditions under which these two definitions of the critical field coincide (the transition being then a sharp phase transition at a precise value). More precisely, we will study the relation of the fields defined above to analogous local ones given purely in terms of spectral data.

First let us recall the result on the asymptotics of the critical field. In \cite{LuPa5,Pan3D}) it was proved that there exists a universal constant $\Theta_0 \approx 0.59$---defined in \eqref{eq:DefTheta0} below---such that
for all (smooth, bounded, simply connected) $\Omega \subset {\mathbb R}^3$ there exists a constant $C>0$ such that for $\kappa$ large
\begin{align}
\label{eq:leading}
\big| H_{C_3}(\kappa) - \frac{\kappa}{\Theta_0} \big| \leq C \kappa^{1/2}.
\end{align}
Here $H_{C_3}(\kappa)$ denotes either $\overline{H}_{C_3}(\kappa)$ or $\underline{H}_{C_3}(\kappa)$.

The local fields are determined by the values where the normal
solution $(0, {\bf F})$ is a {\it not unstable} 
local minimum of ${\mathcal E}_{\kappa, H}$, i.e.
\begin{align}
\overline{H}_{C_3}^{\rm loc}(\kappa) &:=  \inf\{ H>0 \;:\; \text{ for all } H'>H, \Hess {\mathcal E}_{\kappa, H'} \big |_{(0, {\bf F})} \geq 0 \} \;,\nonumber\\
\underline{H}_{C_3}^{\rm loc}(\kappa) &:=  \inf\{ H>0 \;:\;  \Hess {\mathcal E}_{\kappa, H} \big |_{(0, {\bf F})} \geq 0 \}\;.
\end{align}
Since the Hessian, $\Hess {\mathcal E}_{\kappa, H}$, at the normal solution defines the quadratic form
\begin{align}
(\phi,{\bf a}) \mapsto \int_{\Omega} | (-i\nabla + \kappa H {\bf F}) \phi |^2 - \kappa^2 |\phi |^2\,dx
+ (\kappa H)^2 \int_{{\mathbb R}^3} |\curl {\bf a} |^2 \,dx\;, 
\end{align}
we get the equivalent definitions given by 
\begin{align}
\label{eq:SpecLocalDef}
\overline{H}_{C_3}^{\rm loc}(\kappa) &=  \inf\{ H>0 \;:\; \text{ for all } H'>H, \lambda_1(\kappa H') \geq \kappa^2 \} \;, \nonumber \\
\underline{H}_{C_3}^{\rm loc}(\kappa) &=  \inf\{ H>0 \;:\;  \lambda_1(\kappa H) \geq \kappa^2 \}\;.
\end{align}
Here $\lambda_1(B)$ is the the lowest eigenvalue of the magnetic Neumann Laplacian ${\mathcal H}(B)$, i.e. of the self-adjoint operator (with Neumann boundary conditions) associated to the  quadratic form
\begin{align}
\label{eq:Form}
W^{1,2}(\Omega) \ni u \mapsto Q_B(u) :=\int_{\Omega} | p_{B{\bf F}} u |^2\,dx\;.
\end{align}
In other words, ${\mathcal H}(B)$ is the differential operator $p_{B{\bf F}}^2$ with domain
$\{ u \in W^{2,2}(\Omega) \,:\, N\cdot p_{B{\bf F}} u |_{\partial \Omega} = 0\}$.
The operator ${\mathcal H}(B)$ clearly has compact resolvent. \\
The analysis in \cite{LuPa5, Pan3D} implies that \eqref{eq:leading} remains true for the local fields, i.e.
\begin{align}
\label{eq:leadingLoc}
\big| H_{C_3}^{\rm loc}(\kappa) - \frac{\kappa}{\Theta_0} \big| \leq C \kappa^{1/2},
\end{align}
where $H_{C_3}^{\rm loc}(\kappa)$ denotes either $\overline{H}_{C_3}^{\rm loc}(\kappa)$ or $\underline{H}_{C_3}^{\rm loc}(\kappa)$.

Before stating our first main result let us give a third possible definition of the critical fields\footnote{
Our ${\mathcal E}$ is probably the physically most correct of the two. In the literature both ${\mathcal E}$ and ${\mathcal E}^{\rm mod}$ are considered. For example, \cite{LuPa5} study ${\mathcal E}^{\rm mod}$, but in the follow-up paper \cite{Pan3D}, it is ${\mathcal E}$ that is taken as the definition of the functional.}.
In two dimensions the magnetic field energy is usually given as an integral over the domain only. By analogy, one finds the following slight modification of the GL-functional
\begin{multline}
\label{eq:GL_F-modified}
{\mathcal E}^{\rm mod}[\psi,{\bf A}] = {\mathcal
E}^{\rm mod}_{\kappa,H}[\psi,{\bf A}]  =
\int_{\Omega} \Big\{ |p_{\kappa H {\bf A}}\psi|^2 
- \kappa^2|\psi|^2
+\frac{\kappa^2}{2}|\psi|^4\Big\} \,dx\\
+ \kappa^2 H^2 \int_{\Omega}
|\curl {\bf A} - \beta|^2\,dx\;.
\end{multline}
Here $(\psi, {\bf A}) \in W^{1,2}(\Omega;{\mathbb C}) \times W^{1,2}(\Omega;{\mathbb R}^3)$.
Using the gauge invariance of the problem, we can and will assume ${\bf A}$ to be restricted to the smaller space
\begin{align}
\label{eq:LocSpace}
H^1_{\Div}(\Omega) := \big\{ {\bf A} \in W^{1,2}(\Omega;{\mathbb R}^3)
\,\big| \, \Div {\bf A}= 0 \text{ in } \Omega\;,\, N\cdot {\bf A}=0 \text{ on } \partial \Omega \;\big\}\;.
\end{align}
This modified functional leads to a new set of possible values for the critical field.
Let ${\bf F}_{\Omega}$ denote the vector potential in $H^1_{\Div}(\Omega)$ generating the constant magnetic field in $\Omega$, i.e.
\begin{align}
\label{eq:F}
&\left.
\begin{aligned}
\Div {\bf F}_{\Omega} = 0\quad\quad\\
\curl {\bf F}_{\Omega} =\beta \quad\quad
\end{aligned} \right\}
\quad \text{ in } \Omega\;, & N\cdot  {\bf F}_{\Omega}  = 0 \quad \text{ on } \partial \Omega\;.
\end{align}
The new fields are given analogously to the previous ones.
\begin{align}
\overline{H}_{C_3}^{\rm mod}(\kappa) &:=  \inf\{ H>0 \,:\, \text{for all } H'>H, \nonumber\\
&\quad\quad\quad\quad\quad\quad\quad\quad(0, {\bf F}_{\Omega}) \text{ is the unique minimizer of } {\mathcal E}^{\rm mod}_{\kappa,H'}\}\;,\nonumber\\
\underline{H}_{C_3}^{\rm mod}(\kappa) &:= \inf\{ H>0 \,:\, (0, {\bf F}_{\Omega}) \text{ is a minimizer of } {\mathcal E}^{\rm mod}_{\kappa,H}\}\;.
\end{align}
The leading order asymptotics \eqref{eq:leading} also holds for $\overline{H}_{C_3}^{\rm mod}$ and $\underline{H}_{C_3}^{\rm mod}(\kappa)$.

We also state the Euler-Lagrange equations for stationary points of the modified functional
\begin{subequations}
\label{eq:GL-mod}
\begin{align}
\left.\begin{array}{c}
p_{\kappa H {\bf A}}^2\psi =
\kappa^2(1-|\psi|^2)\psi \\
\label{eq:equationA-mod}
\curl^2 {\bf A} =-\tfrac{i}{2\kappa H}(\overline{\psi} \nabla
\psi - \psi \nabla \overline{\psi}) + |\psi|^2 {\bf A}
\end{array}\right\} &\quad \text{ in } \quad \Omega \, ;\\
\left. \begin{array}{c}
(p_{\kappa H {\bf A}} \psi) \cdot N = 0 \\
(\curl {\bf A} - \beta)\times N = 0
\end{array} \right\} &\quad \text{ on } \quad \partial\Omega \, .
\end{align}
\end{subequations}

\begin{remark}~\\
We do not have to define the local fields corresponding to ${\mathcal E}^{\rm mod}$---these coincide with the previously defined local fields. To see this, notice that since $\Omega$ is simply connected and $\curl {\bf F} |_{\Omega} = \curl {\bf F}_{\Omega}=\beta$ there exists a gauge $\phi$ on $\Omega$ such that
$$
{\bf F} |_{\Omega} = {\bf F}_{\Omega} + \nabla \phi\;.
$$
Therefore, the magnetic Neumann operators $(-i\nabla + B {\bf F})^2$ and  $(-i\nabla + B {\bf F}_{\Omega})^2$ on $L^2(\Omega)$ are unitarily equivalent. In particular, 
$$
\inf \Spec (-i\nabla + B {\bf F})^2 = \inf \Spec (-i\nabla + B {\bf F}_{\Omega})^2\;.
$$
\end{remark}

Our first main result (combining Proposition~\ref{prop:Gen_critfields} and Theorem~\ref{thm:critfields}) is that all the critical fields above are contained in the interval $[\underline{H}_{C_3}^{\rm loc}(\kappa) , \overline{H}_{C_3}^{\rm loc}(\kappa) ]$, when $\kappa$ is large. 
We first observe the following general inequalities.

\begin{proposition}
\label{prop:Gen_critfields}~\\
The following general relations hold between the different definitions of $H_{C_3}$:
\begin{align}
\label{eq:lower}
&\underline{H}_{C_3}^{\rm loc}(\kappa) \leq \underline{H}_{C_3}(\kappa) \leq \underline{H}_{C_3}^{\rm mod}(\kappa) \;,\\
\label{eq:upper}
&\overline{H}_{C_3}^{\rm loc}(\kappa) \leq \overline{H}_{C_3}(\kappa) \leq \overline{H}_{C_3}^{\rm mod}(\kappa)\;.
\end{align}
\end{proposition}

For large values of $\kappa$, we have a converse to Proposition~\ref{prop:Gen_critfields}.

\begin{theorem}
\label{thm:critfields}~\\
There exists $\kappa_0>0$ such that for $\kappa \geq \kappa_0$,
\begin{align}
\label{eq:collapse1}&\underline{H}_{C_3}^{\rm loc}(\kappa) = \underline{H}_{C_3}(\kappa) = \underline{H}_{C_3}^{\rm mod}(\kappa) \;,\\
\label{eq:collapse}
&\overline{H}_{C_3}^{\rm loc}(\kappa) = \overline{H}_{C_3}(\kappa) = \overline{H}_{C_3}^{\rm mod}(\kappa)\;.
\end{align}
\end{theorem}

An important consequence of Theorem~\ref{thm:critfields} is that in order to obtain an asymptotic expansion of $H_{C_3}(\kappa)$ for large values of $\kappa$, whichever the definition, one only has to consider the {\it linear} problem of determining $\lambda_1(B)$.

We state the following useful corollary to Theorem~\ref{thm:critfields}.

\begin{cor}
\label{cor:Identical}~\\
Suppose that there exists $B_0>0$ such that $B \mapsto \lambda_1(B)$ is strictly increasing for $B\geq B_0$. Then there exists $\kappa_0 >0$ such that
for $\kappa \in [\kappa_0, \infty)$ one has
\begin{align}
\underline{H}_{C_3}^{\rm loc}(\kappa) = \underline{H}_{C_3}(\kappa) = \underline{H}_{C_3}^{\rm mod}(\kappa)=
\overline{H}_{C_3}^{\rm loc}(\kappa) = \overline{H}_{C_3}(\kappa) = \overline{H}_{C_3}^{\rm mod}(\kappa)\;.
\end{align}
\end{cor}

$\,$

\begin{remark}~\\
The monotonicity of $\lambda_1(B)$ can be proven if one has a sufficiently good asymptotic expansion for large $B$. For domains with smooth boundary (as considered here), we have, $\lambda_1(B)= \Theta_0 B + o(B)$, to leading order. It follows from \cite[Proposition~2.2]{FournaisHelffer2} (the proof is independent of dimension) that if one can prove an asymptotics up to order $o(1)$, i.e.
\begin{align}
\label{eq:Dream}
\lambda_1(B)= \Theta_0 B + \sum_{j=1}^M \alpha_j B^{\gamma_j} + o(1)\;,
\end{align}
with $M \in {\mathbb N}$, $\alpha_j \in {\mathbb R}$, $\gamma_j \in [0,1)$, then $\lambda_1(B)$ is monotonically increasing for large $B$.
However, to the authors' knowledge, there is no example in three
dimensions, where an asymptotics as precise as \eqref{eq:Dream} is
known (except for cylindrical, i.e. effectively two-dimensional cases).
Nevertheless, as we will see below, one can modify the proof of
\cite[Proposition~2.2]{FournaisHelffer2} in order to get the monotonicity of $\lambda_1$ as a consequence of a less demanding asymptotics combined with localization estimates on the ground state eigenfunction.
\end{remark}

The analysis of $H_{C_3}$ described above clearly calls for a clarification of the monotonicity of the function $B \mapsto \lambda_1(B)$. In the two-dimensional situation this has been analyzed in \cite{FournaisHelffer3,FournaisHelffer4}. We will here carry out the similar study of the three-dimensional case.
We will prove (under a generic assumption on the domain $\Omega$) that the mapping $B \mapsto \lambda_1(B)$ is monotonically increasing for sufficiently large values of $B$.

We will work under the following geometric assumptions.
\begin{assumption}\label{ass:Generic1}~\\
The set of boundary points where $\beta$ is tangent to $\partial \Omega$, i.e.
\begin{align}
\Gamma:= \{ x \in \partial \Omega\,\big|\, \beta \cdot N(x) = 0 \},
\end{align}
is a regular submanifold of $\partial \Omega$.
\end{assumption}
 Therefore, $\Gamma$ is a disjoint union of regular curves. We choose an orientation on each such curve, and define the normal curvature at the point $x\in \Gamma$ by
\begin{align}
\label{eq:knonvanish}
k_n(x):= K_x(T(x)\wedge N(x), \beta).
\end{align}
Here $K$ denotes the second fundamental form on $\partial \Omega$, and $T(x)$ is the oriented, unit tangent vector to $\Gamma$ at the point $x$.
We assume that
\begin{assumption}\label{ass:Generic2a}
\begin{align}
\label{eq:genericCurv}
k_n(x) \neq 0, \quad\quad \forall x \in \Gamma.
\end{align}
\end{assumption} and that
\begin{assumption}\label{ass:Generic2b}~\\
The set of points where $\beta$ is tangent to $\Gamma$ is finite.
\end{assumption}

Assumptions~\ref{ass:Generic1}, \ref{ass:Generic2a} and ~\ref{ass:Generic2b} are clearly
generically satisfied. They are for instance satisfied for ellipsoids, whereas a domain containing a cylindrical boundary piece with axis parallel to $\beta$ will violate these assumptions.

We will need the known two-term asymptotics of the ground state energy of ${\mathcal H}(B)$. The following result was proved in \cite{HeMo3} (the corresponding upper bound was also given in \cite{Pan3D} and a less general geometric situation was studied in \cite{HeMo3a}).
\begin{theorem}\label{thm:AsympHeMo}~\\
There exist universal constants $\Theta_0, \widehat{\nu}_0,
\delta_0>0$ (to be defined in \eqref{eq:DefTheta0}, \eqref{eq:delta0},
\eqref{eq:hatnu} below) such that if $\Omega$ satisfies
Assumptions~\ref{ass:Generic1}, ~\ref{ass:Generic2a}  and ~\ref{ass:Generic2b}, then
\begin{align}
\label{eq:asymp}
\lambda_1(B)= \Theta_0 B + \widehat{\gamma}_0 B^{\frac{2}{3}} + {\mathcal O}(B^{\frac{2}{3} - \eta}),
\end{align}
for some $\eta>0$.
\end{theorem}
Here $ \widehat{\gamma}_0$ is defined by
\begin{align}
\widehat{\gamma}_0&:= \inf_{x\in \Gamma} \widetilde{\gamma}_0(x),\\
\label{eq:tildegamma}
\widetilde{\gamma}_0(x)&:= 2^{-2/3} \widehat{\nu}_0 \delta_0^{1/3} 
|k_n(x)|^{2/3}
\Big( \delta_0 + (1-\delta_0)| T(x)\cdot \beta |^2
\Big)^{1/3}\;.
\end{align}

Notice that $\delta_0<1$, so there is no problem with the third root in \eqref{eq:tildegamma}.
Combining this result with Corollary~\ref{cor:Identical}, one gets a two-term asymptotics for $H_{C_3}(\kappa)$:

\begin{cor}\label{cor:asymp}~\\
Suppose that $\Omega\subset {\mathbb R}^3$ satisfies
Assumptions~\ref{ass:Generic1}, ~\ref{ass:Generic2a} and ~\ref{ass:Generic2b}, then one finds
\begin{align}
\label{eq:HC3-HeMo}
H_{C_3}(\kappa) - ( \frac{\kappa}{\Theta_0}- \hat{\gamma}_0 \Theta_0^{-2/3} \kappa^{1/3})=o(\kappa^{1/3})\;,
\end{align}
where $H_{C_3}(\kappa)$ denotes any of the six different (upper or lower) critical fields defined above.
\end{cor}

Corollary~\ref{cor:asymp} is an affirmative answer to a conjecture in
\cite{Pan3D}, however, the conjecture is stated without the geometric
Assumptions~\ref{ass:Generic1},  ~\ref{ass:Generic2a} and  ~\ref{ass:Generic2b}.

\begin{proof}[Proof of Corollary~\ref{cor:asymp}]~\\
By Proposition~\ref{prop:Gen_critfields} and Theorem~\ref{thm:critfields} it suffices to prove that $\underline{H}_{C_3}^{\rm loc}(\kappa)$ and $\overline{H}_{C_3}^{\rm loc}(\kappa)$ have the asymptotics given by \eqref{eq:HC3-HeMo}. But this follows easily from \eqref{eq:asymp}.
\end{proof}

Finally we prove that that $\lambda_1(B)$ is increasing for generic $\Omega$, which implies that the different critical fields coincide, cf. Corollary~\ref{cor:Identical}.

\begin{thm}\label{thm:Derivative}~\\
Let $\Omega \subset {\mathbb R}^3$ satisfy
Assumptions~\ref{ass:Generic1}, ~\ref{ass:Generic2a} and~\ref{ass:Generic2b}. 
Let $\{ \Gamma_1,\ldots, \Gamma_n\}$ be the collection of disjoint smooth curves making up $\Gamma$. 
We assume in addition  that~:
\begin{equation}
\mbox{ For all } j \mbox{ there exists } x \in \Gamma_j
 \mbox{  such that } \widetilde{\gamma}_0(x) > \widehat{\gamma}_0\;.
\end{equation}
Then the directional derivatives
$\lambda_{1,\pm}' := \lim_{\beta\rightarrow 0_{\pm}} \frac{\lambda_1(B+\beta)-\lambda(B)}{\beta}$,
exist and satisfy
\begin{align}
\label{eq:LimitDeriv}
\lim_{B \rightarrow \infty} \lambda_{1,+}'(B) = \lim_{B \rightarrow \infty} \lambda_{1,-}'(B) = \Theta_0\;.
\end{align}
In particular $B \mapsto \lambda_1(B)$ is strictly increasing for $B$ sufficiently large.
\end{thm}

\section{The analysis of $H_{C_3}$}
In this section we give the proof of Theorem~\ref{thm:critfields}. We
aim at giving a simple proof without refering to long technical
papers. Therefore, some of the statements in this section will not be
the best possible ones known in the litterature. In particular, we
avoid the use of the elliptic estimates obtained by `blow-up'
techniques (cf. \cite{Pan3D} and see also \cite{LuPa1,Alm,FournaisHelffer4}).

\subsection{Basic estimates}$\,$\\
In this section we mainly recall a number of results that will be useful in the proof of Theorem~\ref{thm:critfields}.
First we need some weak control of $\lambda_1(B)$.
 
\begin{theorem}[First order eigenvalue bound]
\label{thm:FirstOrder}~\\
Let $\Omega \subset {\mathbb R}^3$ be bounded and simply connected with smooth boundary.
Then
$$
\lambda_1(B) = \Theta_0 B + o(B)\;.
$$
\end{theorem}
Much more precise estimates are proved in \cite{LuPa5,HeMo3,Pan3D} under additional assumptions (cf. Theorem~\ref{thm:AsympHeMo}), but we want to stress that, for the proof of Theorem~\ref{thm:critfields}j, we only need very weak spectral information.

It is a standard consequence (cf. \cite{DGP}) of the maximum principle that a minimizer\footnote{More generally, the inequality \eqref{eq:Linfty} is valid for all stationary points.} $(\psi, {\bf A})$ of ${\mathcal E}_{\kappa, H}$ or ${\mathcal E}^{\rm mod}_{\kappa, H}$ satisfies
\begin{align}
\label{eq:Linfty}
\| \psi \|_{L^\infty} \leq 1\;.
\end{align}
The normalization of our functional ${\mathcal E}_{\kappa, H}$ is such that ${\mathcal E}_{\kappa, H}[0,{\bf F}]=0$. So any minimizer $(\psi, {\bf A})$ will have non-positive energy. Therefore, the only negative term, $-\kappa^2 \| \psi \|_2^2$, in the functional has to control each of the positive terms. This leads to the following basic inequalities for minimizers,
\begin{align}
\label{eq:BasicQuadForm}
&\| p_{\kappa H {\bf A}} \psi \|_2 \leq \kappa \|\psi \|_2, \\
\label{eq:BasicCurl}
&H \| \curl {\bf A} -\beta \|_2 \leq \| \psi \|_2.
\end{align}
Furthermore, using \eqref{eq:Linfty},
\begin{align}
\label{eq:Basic4-2}
\| \psi \|_4^2 \leq  \| \psi \|_2.
\end{align}
The same inequalities remain true for minimizers of ${\mathcal E}^{\rm mod}_{\kappa, H}$.

Finally, we need elliptic estimates for the $\curl$-$\Div$ system:
\begin{theorem}[Ellipticity of the $\curl$-$\Div$ system]
\label{thm:EllipticDivCurl}~\\
There exists a constant $C>0$ such that for all (magnetic fields) ${\bf b} \in L^2({\mathbb R}^3, {\mathbb R}^3)$ with $\Div {\bf b} = 0$, there exists a unique ${\bf a} \in \dot{H}^1({\mathbb R}^3, {\mathbb R}^3)$ such that
$$
\curl {\bf a} = {\bf b}\;, \quad\quad \Div {\bf a} =0\;.
$$
This solution satisfies the estimate
\begin{align}
\label{eq:H1dot}
\| {\bf a} \|_{\dot{H}^1} \leq C \; \| {\bf b} \|_{L^2}\;.
\end{align}
\end{theorem}

\begin{proof}~\\
An argument for this standard result is given in \cite{Giorgi-Phillips}. It is based on the elementary fact that, for $f \in C_0^{\infty}({\mathbb R}^3; {\mathbb R}^3)$ one has
\begin{align}
\label{eq:DivCurl}
\| f \|_{\dot{H}^1} = \int_{{\mathbb R}^3} |\Div f |^2 + |\curl f |^2\,dx\;.
\end{align}
With $\Gamma(x) = \frac{1}{4\pi |x|}$ being the fundamental solution of the Laplacian, the desired solution is (formally)
${\bf a} = - \curl(\Gamma*{\bf b})$.
\end{proof}

\begin{prop}
\label{prop:VectorPotential}~\\
Let $2\leq p \leq 6$ and let $\Omega \subset {\mathbb R}^3$ have bounded measure. Then there exists a constant $C_p>0$ such that for all ${\bf b} \in L^2({\mathbb R}^3, {\mathbb R}^3)$ with $\Div {\bf b} = 0$ the solution ${\bf a}$ given in Theorem~\ref{thm:EllipticDivCurl} satisfies the estimate
\begin{align}
\| {\bf a} \|_{L^p(\Omega)} \leq C_p \; \| {\bf b} \|_{L^2({\mathbb R}^3)}\;.
\end{align}
\end{prop}

\begin{proof}~\\
By \eqref{eq:H1dot} 
and the standard three dimensional Sobolev estimate
\begin{align}
\label{eq:Sobolev}
\| f \|_{L^6({\mathbb R}^3)} \leq C_{\rm Sob}\; \| f \|_{\dot{H}^1} \;,\quad \forall f \in \dot{H}^1({\mathbb R}^3)\;,
\end{align}
the desired estimate holds for $p=6$. Since $\Omega$ has finite measure, H\"{o}lder's inequality implies that $\| {\bf a} \|_{L^p(\Omega)} \leq C \| {\bf a} \|_{L^6(\Omega)}$, for $p\leq 6$. 
\end{proof}

The estimates we actually use below are the ones corresponding to Theorem~\ref{thm:EllipticDivCurl} and Proposition~\ref{prop:VectorPotential} for vector potentials ${\bf a}$ and magnetic fields ${\bf b}$ defined only on $\Omega$.
Recall the definition of $H^1_{\Div}(\Omega)$ in \eqref{eq:LocSpace}.
The next theorem is standard. It is a summary of the results (in the simply connected case) of the discussion in \cite[Appendix 1]{Temam}.

\begin{theorem}
\label{thm:Temam}
~\\
Let $\Omega \subset {\mathbb R}^3$ be smooth, bounded and simply connected and having a finite number of connected boundary components denoted by $S_1, \ldots, S_n$.
Then the space of $L^2$-magnetic fields, i.e. the image of $W^{1,2}(\Omega)$ under the operator $\curl$, satisfies
\begin{align}
{\mathcal B} := \curl W^{1,2}(\Omega) =
\big\{ {\bf b} \in L^2(\Omega) \, \big | \, \Div {\bf b} = 0\;, \; \int_{S_i} {\bf b}\cdot N \,dS =0 \;,\,\forall i \;\big\}\;.
\end{align}
In particular, if $n=1$, we have
\begin{align}
{\mathcal B} = \big\{ {\bf b} \in L^2(\Omega) \, \big | \, \Div {\bf b} = 0 \big\}\;.
\end{align}
Furthermore, for all ${\bf b} \in {\mathcal B}$, there exists a unique ${\bf a} \in H^1_{\Div}(\Omega)$ satisfying
\begin{align}
\label{eq:curlDiv}
\curl {\bf a} = {\bf b}\;. 
\end{align}
 Finally, there exists a constant $C>0$ (depending only on $\Omega$) such that for all ${\bf b} \in {\mathcal B}$, the solution ${\bf a}\in H^1_{\Div}(\Omega)$ to \eqref{eq:curlDiv} satisfies
\begin{align}
\| {\bf a} \|_{W^{1,2}(\Omega)} \leq C\; \| {\bf b} \|_{L^2(\Omega)}\;.
\end{align}
\end{theorem}

We get an analogous estimate to the one in Proposition~\ref{prop:VectorPotential} with unchanged proof.

\begin{prop}
\label{prop:VectorPotentialLocal}~\\
Let $2\leq p \leq 6$ and let $\Omega \subset {\mathbb R}^3$  be bounded and simply connected with smooth connected boundary. Then there exists a constant $C_p>0$ such that for all ${\bf b} \in L^2(\Omega), {\mathbb R}^3)$ with $\Div {\bf b} = 0$ the solution ${\bf a}$ given in Theorem~\ref{thm:Temam} satisfies the estimate
\begin{align}
\| {\bf a} \|_{L^p(\Omega)} \leq C_p \; \| {\bf b} \|_{L^2(\Omega)}\;.
\end{align}
\end{prop}

$\,$

\subsection{Weak decay estimates}~\\
In this subsection we prove the following decay estimate in the
variable normal to the boundary. Much more precise (Agmon type) decay
estimates exist but they depend on a much longer analysis of the
Ginzburg-Landau equations (see \cite{Pan3D}). The estimate below is
very robust and sufficient for our purpose. It extends \cite{BonFo}
who treat the two-dimensional case.

\begin{theorem}[Weak normal decay estimate]
\label{thm:Rough}~\\
Let $\Omega\subset {\mathbb R}^3$ be a bounded domain with Lipschitz boundary.
Then there exist two positive constants $C$ and $C'$, such that, if  
$(\psi, {\bf A})_{\kappa,H}$ is a minimizer of ${\mathcal E}_{\kappa,H}$ or ${\mathcal E}^{\rm mod}_{\kappa, H}$, with 
\begin{align}
\label{eq:AboveC2}
\kappa(H-\kappa) \geq 1/2\,,
\end{align}
then
\begin{align}\label{eq:RoughL2}
\| \psi \|_2^2 \leq C \int_{ \{\sqrt{\kappa(H-\kappa)}\, \dist(x,\partial \Omega)\leq 1 \}} |\psi(x)|^2\,dx 
\leq \frac{C'}{\sqrt{\kappa(H-\kappa)}}.
\end{align}
and
\begin{align}
\label{eq:L-p}
\| \psi \|_{L^2(\Omega)} \leq C_p \;\big[\kappa(H-\kappa)\big]^{-\frac{p-2}{4p}} \| \psi \|_{L^p(\Omega)}\;.
\end{align}
\end{theorem}

\begin{proof}~\\
There is no real modification for the 3D case in comparison with
\cite{BonFo} but we repeat the proof for completeness.
The last inequality in (\ref{eq:RoughL2}) is an easy consequence of \eqref{eq:Linfty},
 since there exists a constant $C_1>0$ (depending only on $\Omega$) such that
\begin{equation}
\meas \{ x\; : \;  \dist(x,\partial \Omega) \leq \lambda\} \leq C_1
\lambda\;,\;
 \forall \lambda \in (0,2]\;.
\end{equation}

Let $\chi \in C^{\infty}({\mathbb R})$ be a standard non-decreasing cut-off function,
$$
\chi = 1 \quad \text{ on } [1,\infty), \quad\quad \chi = 0 \quad \text{ on } (-\infty,1/2).
$$
Define $\lambda:={1}/{\sqrt{\kappa(H-\kappa)}}$ and $\chi_{\lambda}:\Omega \rightarrow {\mathbb R}$ by
$$
\chi_{\lambda}(x)  := \chi(\dist(x,\partial \Omega)/\lambda).
$$
Then $\chi_{\lambda}$ is a Lipschitz function and $\supp \chi_{\lambda} \subset \Omega$. Combining the standard localization formula and \eqref{eq:equationA}, we find
\begin{align}
\label{eq:LocFirst}
\int_{\Omega} | p_{\kappa H {\bf A}} (\chi_{\lambda} \psi) |^2 \,dx-
\int_{\Omega} |\nabla \chi_{\lambda}|^2 |\psi |^2\,dx
&=
\Re \langle \chi_{\lambda}^2 \psi,  {\mathcal H}_{\kappa H {\bf A}} \psi \rangle \nonumber\\
&=
\kappa^2 \int |\chi_{\lambda} \psi|^2\,dx - 
\kappa^2 \int \chi_{\lambda}^2 |\psi|^4\,dx.
\end{align}
Note also that, by integration by parts, one has, since $\chi_{\lambda} \psi$ has compact support,  the following important though elementary inequality,
\begin{align}\label{eq:CompSuppBa}
\int_{\Omega} | p_{\kappa H {\bf A}} (\chi_{\lambda} \psi) |^2\,dx
&\geq
\kappa H \int_{\Omega} (\curl {\bf A})_3 |\chi_{\lambda}\psi |^2 \,
\end{align}
Then, using  \eqref{eq:BasicCurl}, we get
\begin{align}\label{eq:CompSuppBb}
\int_{\Omega} | p_{\kappa H {\bf A}} (\chi_{\lambda} \psi) |^2\,dx
&\geq
\kappa H \|  \chi_{\lambda} \psi \|_2^2  - 
\kappa H \| \curl {\bf A} -\beta \|_2 \|  \chi_{\lambda} \psi \|_4^2\nonumber\\
&\geq \kappa H \|  \chi_{\lambda} \psi \|_2^2 - \frac{1}{4}\| \psi\|_2^2 - \kappa^2 \|\chi_{\lambda} \psi \|_4^4.
\end{align}
Using \eqref{eq:Basic4-2} and \eqref{eq:BasicCurl}, we get from \eqref{eq:LocFirst} and \eqref{eq:CompSuppBb} that
\begin{align*}
&\kappa (H- \kappa) \|  \chi_{\lambda} \psi \|_2^2\\
&\leq
\frac{1}{4} \| \psi \|_2^2 + \| \chi' \|_{\infty}^2 \lambda^{-2} \int_{\{\dist(x,\partial \Omega) \leq \lambda\}}
|\psi(x)|^2\,dx  + \kappa^2 \int (\chi_{\lambda}^4- \chi_{\lambda}^2) |\psi|^4\,dx.
\end{align*}
Notice that the last integral is negative and we thus find by 
splitting $\| \psi \|_2^2$
\begin{align*}
\{\kappa (H- \kappa) -&1/4 \} \|  \chi_{\lambda} \psi \|_2^2\\
&\leq (\| \chi' \|_{\infty}^2 \lambda^{-2} +1/4) \int_{\{\dist(x,\partial \Omega) \leq \lambda\}}
|\psi(x)|^2\,dx.
\end{align*}
By assumption 
$$ \kappa(H-\kappa) -1/4 \geq \kappa(H-\kappa)/2\;.
$$
Moreover
 the conditions on $\chi$ and  $\kappa(H-\kappa)$ imply that
$$
 \| \chi' \|_{\infty}^2 \lambda^{-2} +1/4 \leq \left(\| \chi'
  \|_{\infty}^2+1\right)  \lambda^{-2}\;.$$
Thus,
\begin{align}
\|  \chi_{\lambda} \psi \|_2^2
\leq 2( \| \chi' \|_{\infty}^2 +1)\int_{\{\dist(x,\partial \Omega) \leq \lambda\}}
|\psi(x)|^2\,dx.
\end{align}
Consequently,
\begin{align}
\|  \psi \|_2^2
\leq 4 \left(\| \chi' \|_{\infty}^2 + 1\right)\int_{\{\dist(x,\partial \Omega) \leq \lambda\}}
|\psi(x)|^2\,dx.
\end{align}
This finishes the proof of \eqref{eq:RoughL2}.

The $L^p$ estimate \eqref{eq:L-p} is a consequence of the first inequality in \eqref{eq:RoughL2} and H\"{o}lder's inequality.
\end{proof}
$\,$
\subsection{Equal fields}~\\
We first give the easy proof of Proposition~\ref{prop:Gen_critfields}.

\begin{proof}[Proof of Proposition~\ref{prop:Gen_critfields}]~\\
{\it The inequality }$\underline{H}_{C_3}^{\rm loc}(\kappa) \leq \underline{H}_{C_3}(\kappa) $:\\
Suppose $H < \underline{H}_{C_3}^{\rm loc}(\kappa)$. Then $\lambda_1(\kappa H) < \kappa^2$. Let $\psi$ be a ground state for ${\mathcal H}(\kappa H)$. 
We use, for $\eta>0$, $(\eta \psi, {\bf F})$ as a trial state in ${\mathcal E}_{\kappa, H}$,
\begin{align*}
{\mathcal E}_{\kappa, H}[\eta \psi, {\bf F}] = (\lambda_1(\kappa H) - \kappa^2) \eta^2 \| \psi \|_{L^2(\Omega)}^2 + \frac{\kappa^2}{2} \eta^4 \| \psi \|_{L^4(\Omega)}^4\;.
\end{align*}
Since $\lambda_1(\kappa H) - \kappa^2<0$, we get
${\mathcal E}_{\kappa, H}[\eta \psi,{\bf F}] <0$ for $\eta$ sufficiently small.
Thus $(0, {\bf F})$ is not a minimizer for ${\mathcal E}_{\kappa, H}$. Since $H < \underline{H}_{C_3}^{\rm loc}(\kappa)$ was arbitrary, this proves that $\underline{H}_{C_3}^{\rm loc}(\kappa) \leq \underline{H}_{C_3}(\kappa)$.\\
{\it The inequality} $\underline{H}_{C_3}(\kappa) \leq \underline{H}_{C_3}^{\rm mod}(\kappa)$:\\
Suppose $H < \underline{H}_{C_3}(\kappa)$. Then there exists $(\psi, {\bf A}) \in W^{1,2}(\Omega) \times \dot{H}^1_{{\bf F}, \Div}$ such that ${\mathcal E}_{\kappa, H}[\psi, {\bf A}] < 0$. By restriction ${\bf A}$ defines an element $\tilde{A} \in W^{1,2}(\Omega, {\mathbb R}^3)$ and we get the following simple inequalities,
\begin{align}
{\mathcal E}^{\rm mod}_{\kappa, H}[\psi, \tilde{A}] \leq {\mathcal E}_{\kappa, H}[\psi, {\bf A}] < 0\;.
\end{align}
Thus $(0, {\bf F}_{\Omega})$ is not a minimizer for ${\mathcal E}^{\rm mod}_{\kappa, H}$. 
Since $H < \underline{H}_{C_3}(\kappa)$ was arbitrary, this proves that $\underline{H}_{C_3}(\kappa) \leq \underline{H}_{C_3}^{\rm mod}(\kappa)$.
This finishes the proof of \eqref{eq:lower}.\\
{\it The inequality} $\overline{H}_{C_3}^{\rm loc}(\kappa) \leq \overline{H}_{C_3}(\kappa) $:\\ 
Suppose $H>\overline{H}_{C_3}(\kappa)$. Then $(0,{\bf F})$ is the only minimizer of ${\mathcal E}_{\kappa, H}$. In particular, for all $s \in {\mathbb R}$ and all $\phi, {\bf A}$,
$$
{\mathcal E}_{\kappa, H}[s\phi,{\bf F} + s {\bf A}]
\geq {\mathcal E}_{\kappa, H}[0,{\bf F}] = 0\;.
$$
This implies that $\Hess {\mathcal E}_{\kappa, H}\big |_{(0, {\bf F})} \geq 0$.
Since $H>\overline{H}_{C_3}(\kappa)$ was arbitrary, this proves that $\overline{H}_{C_3}^{\rm loc}(\kappa) \leq \overline{H}_{C_3}(\kappa)$.\\
{\it The inequality} $\overline{H}_{C_3}(\kappa) \leq \overline{H}_{C_3}^{\rm mod}(\kappa) $:\\ 
Let $H>\overline{H}_{C_3}^{\rm mod}(\kappa)$ and let $(\psi, {\bf A}) \in W^{1,2}(\Omega) \times \dot{H}^1_{{\bf F}, \Div}$ be a minimizer of ${\mathcal E}_{\kappa,H}$. Then, since ${\mathcal E}_{\kappa,H}[0,{\bf F}]=0$, we find, with $\tilde{A}$ being the restriction of ${\bf A}$ to $\Omega$,
$$
0 \leq {\mathcal E}^{\rm mod}_{\kappa,H}[\psi,\tilde{A}] \leq {\mathcal E}_{\kappa,H}[\psi,{\bf A}]\leq 0\;.
$$
Furthermore, since $H>\overline{H}_{C_3}^{\rm mod}(\kappa)$, the first inequality can only be an equality if $\psi =0$. But then we find that 
$$
0 = {\mathcal E}_{\kappa,H}[\psi,{\bf A}] = (\kappa H)^2 \int_{{\mathbb R}^3} |\curl {\bf A} - \beta |^2\,dx,
$$
i.e. $\curl {\bf A} = \beta$, which implies that ${\bf A}$ is gauge equivalent to ${\bf F}$. Thus, for all $H>\overline{H}_{C_3}^{\rm mod}(\kappa)$ the only minimizer of ${\mathcal E}_{\kappa,H}$ is the normal state $(0,{\bf F})$. Therefore we have proved the inequality $\overline{H}_{C_3}(\kappa) \leq \overline{H}_{C_3}^{\rm mod}(\kappa)$.
This finishes the proof of \eqref{eq:upper}.
\end{proof}

The proof of \eqref{eq:collapse} is essentially identical to the proof of the corresponding statement in two dimensions given in \cite{FournaisHelffer2}. The small difference occurs in the use of Proposition \ref{prop:VectorPotentialLocal}. We give the proof here for completeness
 and to emphasize that it is remarkable that this argument is sufficiently robust to be generalized to the three dimensional setting where the spectral asymptotics is much more complicated than for planar regions.

Define the sets
\begin{align}
&{\mathcal N}(\kappa) := \{ H \in {\mathbb R} \,\big|\, {\mathcal E}_{\kappa,H} \text{ has a non-trivial minimizer} \}\;, \\
&{\mathcal N}^{\rm mod}(\kappa) := \{ H \in {\mathbb R} \,\big|\, {\mathcal E}^{\rm mod}_{\kappa,H} \text{ has a non-trivial minimizer} \}\;, \\
&{\mathcal N}_{\rm loc}(\kappa) := \{ H \in {\mathbb R} \,\big|\, \lambda_1(\kappa H) < \kappa^2 \}\;.
\end{align}
By evaluating the functional ${\mathcal E}_{\kappa,H}$ (resp ${\mathcal E}^{\rm mod}_{\kappa,H}$) in the state $(\eta \psi, {\bf F})$ (resp $(\eta \psi, {\bf F}_{\Omega})$), with $\eta$ small and $\psi$ being the ground state of ${\mathcal H}(\kappa H)$, one gets the inclusions,
\begin{align}
\label{eq:EasyIncl}
{\mathcal N}_{\rm loc}(\kappa) \subseteq {\mathcal N}(\kappa)\;,\quad
{\mathcal N}_{\rm loc}(\kappa) \subseteq {\mathcal N}^{\rm mod}(\kappa)\;.
\end{align}
This, of course, is analogous to Proposition~\ref{prop:Gen_critfields}. We will prove that the converse inclusion holds for large $\kappa$, so we get
\begin{thm}
\label{thm:InclusionSets}~\\
Let $\Omega$ be bounded and simply-connected with smooth boundary. Then there exists $\kappa_0>0$ such that for all $\kappa \geq \kappa_0$ we have
\begin{align}
\label{eq:DiffIncl}
{\mathcal N}_{\rm loc}(\kappa) = {\mathcal N}(\kappa)={\mathcal N}^{\rm mod}(\kappa)\;.
\end{align}
\end{thm}

Of course, \eqref{eq:EasyIncl} implies that we only need to prove inclusions in one direction in \eqref{eq:DiffIncl}. 
We will only prove $ {\mathcal N}^{\rm mod}(\kappa) \subseteq
{\mathcal N}_{\rm loc}(\kappa)$,
 the similar inclusion for ${\mathcal N}(\kappa)$ being proven in exactly the same way.

\begin{lemma}
\label{lem:Important}~\\
Let $c>0$. Then there exists $\kappa_0>0$ such that if 
\begin{align}
H - \kappa \geq c \kappa,
\end{align}
and $(\psi, {\bf A})_{\kappa, H}$ is a nontrivial minimizer of ${\mathcal E}^{\rm mod}_{\kappa, H}$ with
$\kappa\geq \kappa_0$, then
\begin{align}
\label{eq:LoEn}
\kappa^2 - \lambda_1(\kappa H) > 0\;.
\end{align}
Actually, we have the more precise estimate
\begin{align}
\label{eq:Precise}
0 < \frac{ \kappa^2 \| \psi \|_2^2 -  Q_{\kappa H {\bf A}}[\psi]}{\| \psi \|_2^2}
\leq \big(1 +{\mathcal O}(\kappa^{-\frac{1}{4}})\big) (\kappa^2 - \lambda_1(\kappa H))\;.
\end{align}
\end{lemma}

\begin{proof}~\\
Recall the rough asymptotic estimate
\begin{align}
\label{eq:rough}
\big| H_{C_3}^{\rm mod}(\kappa) - \frac{\kappa}{\Theta_0} \big| \leq C \sqrt{\kappa}\;,
\end{align}
stated in the introduction for any definition of $H_{C_3}$.

By definition of $\overline{H}^{\rm mod}_{C_3}(\kappa)$, nontrivial minimizers only exist below $\overline{H}^{\rm mod}_{C_3}(\kappa)$, so we may assume that
$$
(1+c)\kappa \leq H \leq \overline{H}^{\rm mod}_{C_3}(\kappa) \;.
$$
Since $(\psi, {\bf A})$ is non-trivial, we get that~:
\begin{align}
\kappa^2 \| \psi \|_2^2 >  Q_{\kappa H {\bf A}}[\psi]\;.
\end{align}
We define 
\begin{align}
\Delta &:= \kappa^2 \| \psi \|_2^2 - Q_{\kappa H {\bf A}}[\psi]\;.
\end{align}
Notice, that
the GL-equation gives
\begin{align}
\label{eq:4-new}
 \| \psi\|_4^4 = \frac{\Delta}{\kappa^2}\;.
\end{align}

We are in a situation where Theorem~\ref{thm:Rough}, can be applied. 
Therefore, we get by \eqref{eq:L-p} with $p=4$
\begin{align}
\label{eq:BeforeL2-new}
\| \psi \|_2 \leq C \kappa^{-\frac 14}\| \psi \|_4\;.
\end{align}
Coming back to \eqref{eq:4-new}, we get
\begin{align}
\label{eq:L2-new}
\| \psi \|_2 \leq C \kappa^{-\frac{3}{4}} \Delta^{\frac{1}{4}}\;.
\end{align}
We now replace ${\bf A}$ by ${\bf F}_{\Omega}$ in the quadratic form. Denote ${\bf a}={\bf A}-{\bf F}_{\Omega}$ and ${\bf b}=\curl ({\bf A} -{\bf F}_{\Omega})$. Then the Cauchy-Schwarz inequality implies that 
\begin{align}
\label{eq:GaarDet-new}
0 < \Delta \leq \big[ \kappa^2 - (1-\rho) \lambda_1(\kappa H) \big] \| \psi \|_2^2
+ \rho^{-1} (\kappa H)^2 \int_{\Omega} | {\bf a} \psi |^2 \,dx\;,
\end{align}
for all $0<\rho$.\\
Proposition~\ref{prop:VectorPotentialLocal} implies 
$$
\|{\bf a}\|_{L^4(\Omega)}  \leq C \| {\bf b} \|_{L^2(\Omega)}\;,
$$
Therefore,
\begin{align}
\label{eq:NewaNew}
(\kappa H)^2 \|{\bf a}\|_4^2 \leq C (\kappa H)^2 \| \curl A -1 \|_2^2 \leq C \Delta\;.
\end{align}
Here we used that ${\mathcal E}^{\rm mod}_{\kappa,H}[\psi,{\bf A}] \leq 0$ to get the last estimate.

We now insert \eqref{eq:NewaNew}, \eqref{eq:4-new}, and \eqref{eq:L2-new} in \eqref{eq:GaarDet-new}.
\begin{align}
0 < \Delta &\leq \big[ \kappa^2 - (1-\rho) \lambda_1(\kappa H) \big] \| \psi \|_2^2
+ \rho^{-1} (\kappa H)^2 \| {\bf a} \|_4^2 \|\psi \|_4^2 \nonumber\\
&\leq \big[ \kappa^2 - \lambda_1(\kappa H) \big] \| \psi \|_2^2
+ C \rho \lambda_1(\kappa H) \Delta^{\frac{1}{2}} \kappa^{-\frac{3}{2}}
+ C \rho^{-1} \Delta \frac{\sqrt{\Delta}}{\kappa}
\end{align}
Upon choosing
$\rho = \sqrt{\Delta} \kappa^{-\frac{3}{4}}$, and using that
$\lambda_1(\kappa H) < C \kappa^2$, we find, for another constant
$\widetilde C$,
\begin{align}
0 < \Delta &\leq \big[ \kappa^2 - \lambda_1(\kappa H) \big] \| \psi \|_2^2
+ \widetilde C  \Delta \kappa^{-\frac{1}{4}}\;.
\end{align}
 When $\kappa$ is so big that $\widetilde C \kappa^{-\frac{1}{4}} <1$, we therefore get
\begin{align} \label{inequalitya}
0 < (1- \widetilde C \kappa^{-\frac{1}{4}}) \Delta 
\leq \big[ \kappa^2 - \lambda_1(\kappa H) \big] \| \psi \|_2^2\;.
\end{align}
Since $\psi$ cannot vanish identically for a non-trivial minimizer, this shows both \eqref{eq:LoEn} and \eqref{eq:Precise}.
\end{proof}

\section{Monotonicity of $\lambda_1$}
\subsection{Spectral theory}$\,$\\
In this subsection we will recall a few results of the spectral theory of
some important model operators. This will permit to define precisely
the constants $\Theta_0$, $\delta_0$
 and $\hat \nu_0$ appearing in the main statements.

$\,$
\subsubsection{The model on ${\mathbb R}^2_+$}~\\
Consider first the self-adjoint operator $P_{{\mathbb R}^2_+}$ defined on $L^2({\mathbb R}^2_+)$ (with ${\mathbb R}^2_+:= \{ (s,t) \in {\mathbb R}^2\,: \, t>0\}$) by the quadratic form,
$$
\{ u \in L^2({\mathbb R}^2_+)\,:\, (-i\nabla + {\bf F}_{{\mathbb R}^2})u \in L^2({\mathbb R}^2_+)\} \ni u
\mapsto \int_{{\mathbb R}^2_+} \big| (-i\nabla + {\bf F}_{{\mathbb R}^2})u \big|^2\,ds dt .
$$
Here ${\bf F}_{{\mathbb R}^2}(s,t) := \frac{1}{2}(-t,s)$.

The first constant that we need to define is
\begin{align}
\label{eq:DefTheta0}
\Theta_0 := \inf \Spec P_{{\mathbb R}^2_+}.
\end{align}
It is known that $\frac{1}{2} < \Theta_0 < 1$.\\
A unitary transform reduces the study of $P_{{\mathbb R}^2_+}$ to the study of the Neumann realization of $D_t^2 + (t-s)^2$ on $L^2({\mathbb R}^2_+)$, which again reduces to the study of the family of operators ($s \in {\mathbb R}$), 
$$
{\mathfrak h}(s) := D_t^2 + (t-s)^2,
$$
on $L^2({\mathbb R}_+)$ with Neumann boundary conditions. In particular, with the notation $\mu(s) := \inf \Spec {\mathfrak h}(s)$, we have $\Theta_0 = \inf_{s \in {\mathbb R}} \mu(s)$. One can prove that this infimum is a unique, non-degenerate minimum. I.e. there exists a unique $s_0 \in {\mathbb R}$ such that
$\Theta_0 = \mu(s_0)$, and 
\begin{align}
\label{eq:delta0}
\delta_0 := \frac{\mu''(s_0)}{2}> 0.
\end{align}
Actually, $\delta_0 \in (0,1)$, and an alternative expression for it was derived in \cite{FournaisHelffer1}.
For more information on the spectral theory of $P_{{\mathbb R}^2_+}$ see \cite{DaHe,HeMo1}.

$\,$
\subsubsection{The model on ${\mathbb R}^3_{+}$}~\\
The next model operator to study is the case $\Omega = {\mathbb R}^3_{+}:=\{ (x_1,x_2,x_3) \in {\mathbb R}^3 \,:\, x_1>0\}$, $\beta = (\sin \theta, \cos \theta, 0)$, $B=1$ in \eqref{eq:Form}.
In other words, and with a particular gauge choice, it is the Neumann realization of the differential operator
\begin{align}
P(\beta_1,\beta_2) :=D_{x_1}^2 + D_{x_2}^2+(D_{x_3}+ \beta_2 x_1 - \beta_1 x_2 )^2,
\end{align}
in $L^2({\mathbb R}^3_{+})$.
We define
\begin{align}
\label{eq:Defsigma}
\sigma(\theta) := \inf \Spec P(\sin \theta, \cos \theta).
\end{align}
We will need a number of properties of this spectral function. For proofs and more precise results see \cite{HeMo2,LuPa5}.

\begin{prop}~\\
The function $\sigma(\theta)$ defined on $[-\pi/2, \pi/2]$ is continuous and even and satisfies
\begin{itemize}
\item $\sigma$ is strictly increasing on $[0, \pi/2]$.
\item $\sigma(0) = \Theta_0$ and $\sigma(\pi/2) = 1$.
\item We have the following expansion for small $\theta$:
\begin{align}
\label{eq:SmallTheta}
\sigma(\theta) = \Theta_0 + \alpha_1 | \theta| + {\mathcal O}(\theta^2),
\end{align}
where $\alpha_1=\sqrt{\delta_0}>0$ (with $\delta_0$ from \eqref{eq:delta0}).
\end{itemize}
\end{prop}
$\,$
\subsubsection{Montgomery's model}~\\
We finally mention briefly a model on $L^2({\mathbb R}^2)$ which has been studied in \cite{Mon,HeMo0,PanKwek,He3}. This model has magnetic field vanishing on a line. Define $P_{\rm M}$ to be the self-adjoint realization of
$$
P_{\rm M} :=D_x^2 + (D_y -x^2)^2,
$$
on $L^2({\mathbb R}^2)$.
Define the constant $\widehat{\nu}_0$ as
\begin{align}
\label{eq:hatnu}
\widehat{\nu}_0 := \inf \Spec P_{\rm M}.
\end{align}

\subsection{Localization estimates}~\\
We start by recalling the decay in the direction normal to the boundary.
We will often use the notation 
\begin{align}
t(x):=\dist(x, \partial \Omega).
\end{align}
Now, if $\phi \in C_0^{\infty}(\Omega)$, i.e. has support away from
the boundary, a simple standard integration by parts (already used in 
 a very similar way in \eqref{eq:CompSuppBa})
 implies that
\begin{align}
\label{eq:EnergyInterior}
Q_B(\phi) \geq B \| \phi \|_2^2.
\end{align}
It is a consequence of this elementary inequality (and the fact that $\Theta_0 < 1$) that ground states are exponentially localized near the boundary. We give the result without proof since this has been proven by many authors in this and similar situations.

\begin{thm}\label{thm:AgmonNormal}~\\
Let $\Omega \subset {\mathbb R}^3$ be a bounded open set with smooth
boundary. Then there exist constants $C, a_1 >0, B_0$ such that
\begin{align}
\int_{\Omega} e^{2a_1 B^{1/2} t(x)} \Big( |\psi_B(x)|^2 + B^{-1} |(-i\nabla + B {\bf F})\psi_B(x)|^2\Big)\,dx \leq C \| \psi_B\|_2^2,
\end{align}
for all $B\geq B_0$, and all ground states  $\psi_B$ of the operator ${\mathcal H}(B)$.
\end{thm}

We will mainly use this localization result in the following form.
\begin{cor}\label{cor:NormalAgmon}~\\
Let $\Omega \subset {\mathbb R}^3$ be a bounded open set with smooth boundary. Then for all $n \in {\mathbb N}$ there exists $C_n>0$ such that
$$
\int t(x)^n |\psi_B(x)|^2\,dx \leq C_n B^{-n/2} \| \psi_B \|_2^2,
$$
for all $B>0$ and all ground states $\psi_B$ of the operator ${\mathcal H}(B)$.
\end{cor}

We now define tubular neighborhoods of the boundary as follows. For $\epsilon>0$, define
\begin{align}
B(\partial\Omega,\epsilon) := \{x \in \Omega \,:\, t(x) \leq \epsilon\}.
\end{align}
For sufficiently small $\epsilon_0$ we have that for all $x \in B(\partial\Omega,2\epsilon_0)$ exists a unique point $y=y(x) \in \partial \Omega$ such that $t(x) = \dist(x,y)$.
We fix such an $\epsilon_0$ in the rest of the paper.

Define, for $y \in \partial \Omega$, the function $\theta(y) \in [-\pi/2, \pi/2]$ by
\begin{align}
\sin \theta(y) := - \beta\cdot N(y).
\end{align}
We extend $\theta$ to the tubular neighborhood $B(\partial\Omega,2\epsilon_0)$ by
$\theta(x) := \theta( y(x))$.

In order to obtain localization estimates in the (boundary) variable normal to $\Gamma$, we use the following operator inequality (see \cite[Theorem 4.3]{HeMo3}).

\begin{thm}\label{thm:WB}~\\
Let $\Omega \subset {\mathbb R}^3$ be a bounded open set with smooth boundary.
Let $B_0$ be chosen such that $B_0^{-3/8} = \epsilon_0$ and 
define, for $B\geq B_0, C>0$ and $x \in \Omega$,
\begin{align}
W_{B,C}(x) :=\begin{cases} B - C B^{3/4}, & t(x) \geq 2 B^{-3/8}, \\
B \sigma(\theta(x)) - C B^{3/4}, & t(x) < 2 B^{-3/8}.
\end{cases}
\end{align}
Then, there exists $C_0$ such that  
\begin{align}
{\mathcal H}(B) \geq W_{B,C},
\end{align}
(in the sense of quadratic forms) for all $B\geq B_0$ and all $C\geq C_0$.
\end{thm}

We use Theorem~\ref{thm:WB} to prove Agmon type estimates on the boundary.

\begin{thm}\label{thm:TangentAgmon}~\\
Suppose that $\Omega\subset {\mathbb R}^3$ satisfies
 Assumptions~\ref{ass:Generic1},~\ref{ass:Generic2a} and 
 ~\ref{ass:Generic2b}. Define for $x \in \partial \Omega$, 
$$
d_{\Gamma}(x) := \dist_{\pa \Omega}(x, \Gamma),
$$ 
where $\dist_{\pa \Omega}$ is the distance associated to the induced
metric
 of $\pa \Omega$,  
and extend $d_{\Gamma}$ to a tubular neighborhood of the boundary by $d_{\Gamma}(x):= d_{\Gamma}(y(x))$, where $y(x)$ is the unique boundary point closest to $x$.\\
Then there exist constants $C, a_2>0, B_0$, such that
\begin{align}
\label{eq:TangentAgmon}
\int_{B(\partial \Omega, \epsilon_0)} e^{2 a_2 B^{1/2} d_{\Gamma}(x)^{3/2}} |\psi_B(x)|^2\,dx
\leq C e^{ C B^{1/8}} \| \psi_B\|_2^2,
\end{align}
for all $B\geq B_0$ and all ground states $\psi_B$ of ${\mathcal H}(B)$.
\end{thm}

It is useful to collect the following easy consequence\footnote{ It is
  enough
 to consider separately the cases when $d_\Gamma \leq E B^{-\frac 14}$
 and the case when $d_\Gamma \geq E B^{-\frac 14}$
 with $E$ large enough.} .

\begin{cor}\label{cor:AgmonBdry}~\\
Suppose that $\Omega\subset {\mathbb R}^3$ satisfies
 Assumptions~\ref{ass:Generic1}, ~\ref{ass:Generic2a} and ~\ref{ass:Generic2b}.
Then for all $n \in {\mathbb N}$ there exists $C_n>0$ such that
\begin{align}
\int_{B(\partial \Omega, \epsilon_0)} d_{\Gamma}(x)^n |\psi_B(x)|^2\,dx
\leq C_n B^{-n/4} \| \psi_B \|_2^2,
\end{align}
for all $B>0$ and all ground states $\psi_B$ of ${\mathcal H}(B)$.
\end{cor}

$\,$

\begin{proof}[Proof of Theorem~\ref{thm:TangentAgmon}]~\\
We may clearly assume that $\| \psi_B \|_2 =1$. 
Let $\chi_1, \chi_2 \in C^{\infty}({\mathbb R})$ such that
$\chi_1$ is decreasing, $\chi_1 \equiv 0$ on $[2, +\infty)$, $\chi_1 \equiv 1$ on $(-\infty, 1]$,
$\chi_2$ is increasing, $\chi_1 \equiv 1$ on $[2, +\infty)$, $\chi_1 \equiv 0$ on $(-\infty, 1]$.

By the standard localization formula we find, since ${\mathcal H}(B) \psi_B = \lambda_1(B) \psi_B$,
\begin{align}
\label{eq:IMS}
\lambda_1(B) &\Big\| \chi_1(\frac{t}{\epsilon_0}) \chi_2(D B^{1/4}
 d_{\Gamma})
 e^{a_2 B^{1/2} d_{\Gamma}^{3/2}} \psi_B \Big\|_2^2 \nonumber\\
&=
Q_B\Big[  \chi_1(\frac{t}{\epsilon_0}) \chi_2(DB^{1/4} d_{\Gamma})
 e^{a_2 B^{1/2} d_{\Gamma}^{3/2}} \psi_B\Big]\nonumber\\
&\quad\quad-
\int \Big| \nabla\big( \chi_1(\frac{t}{\epsilon_0}) \chi_2(D B^{1/4}
 d_{\Gamma})
 e^{a_2 B^{1/2} d_{\Gamma}^{3/2}}\big)\Big|^2 \,
| \psi_B|^2\,dx\,.
\end{align}
Here $D$ is a sufficiently small constant to be determined.\\
We estimate using Theorem~\ref{thm:WB},
\begin{multline}
\label{eq:UsingWB}
Q_B\Big[  \chi_1(\frac{t}{\epsilon_0}) \chi_2(DB^{1/4} d_{\Gamma}) e^{a_2 B^{1/2} d_{\Gamma}^{3/2}} \psi_B\Big] \\
\geq 
\int W_B(x) \Big|\chi_1(\frac{t}{\epsilon_0}) \chi_2(D B^{1/4}
 d_{\Gamma}) e^{a_2 B^{1/2} d_{\Gamma}^{3/2}} \psi_B \Big|^2\,dx .
\end{multline}
Also,
\begin{align}
\label{eq:Grad2}
\Big| \nabla\big( \chi_1(\frac{t}{\epsilon_0}) \chi_2(DB^{1/4}
 d_{\Gamma})
 &e^{a_2 B^{1/2} d_{\Gamma}^{3/2}}\big)\Big|^2\nonumber\\
&\leq
2 \Big| \nabla \chi_1(\frac{t}{\epsilon_0})\Big|^2 \chi_2^2(DB^{1/4} 
d_{\Gamma}) e^{2a_2 B^{1/2} d_{\Gamma}^{3/2}}\nonumber\\
&\quad+
2 \big| \nabla \chi_2(D B^{1/4} d_{\Gamma})\big|^2 
\chi_1^2(\frac{t}{\epsilon_0}) e^{2a_2 B^{1/2} d_{\Gamma}^{3/2}}\nonumber\\
&\quad+
 \frac{9}{2} a_2^2 B d_{\Gamma} \chi_1^2(\frac{t}{\epsilon_0})
 \chi_2^2(D B^{1/4} d_{\Gamma})
e^{2a_2 B^{1/2} d_{\Gamma}^{3/2}}.
\end{align}
Combining, \eqref{eq:IMS}, \eqref{eq:UsingWB} and \eqref{eq:Grad2}, we find
\begin{align}
\label{eq:AlmostDone}
\int\big( &W_B(x) - \lambda_1(B) -  \frac{9}{2} a_2^2 B
 d_{\Gamma}(x)\big)
 \chi_1^2(\frac{t}{\epsilon_0}) \chi_2^2(D B^{1/4} d_{\Gamma})
e^{2a_2 B^{1/2} d_{\Gamma}^{3/2}} |\psi_B|^2\,dx\nonumber\\
&\leq
C \int_{B(\partial\Omega,2\epsilon_0)\setminus
 B(\partial\Omega,\epsilon_0)}
 e^{2a_2 B^{1/2} d_{\Gamma}^{3/2}} |\psi_B|^2\,dx \nonumber\\
&\quad
+ C B^{1/2} e^{4 \sqrt{2}a_2 D^{-\frac 32} B^\frac 18} \int_{\{ x \in
 B(\partial\Omega,2\epsilon_0)\,:
 \, B^{1/4} d_{\Gamma}(x) \leq \frac 2D \}} |\psi_B(x)|^2\,dx.
\end{align}
Since $\Omega$ is bounded there exists $D_1>0$ such that $d_{\Gamma}(x) \leq D_1$ for all $x$. Thus we can estimate, with $a_1$ being the constant from Theorem~\ref{thm:AgmonNormal},
\begin{align}
\label{eq:EstNormal}
\int_{B(\partial\Omega,2\epsilon_0)\setminus
 B(\partial\Omega,\epsilon_0)}
 &e^{2a_2 B^{1/2} d_{\Gamma}^{3/2}} |\psi_B|^2\,dx\nonumber\\
&\leq e^{2a_2 B^{1/2} D_1^{3/2}} e^{-2a_1 B^{1/2} \epsilon_0}
\int_{\Omega} e^{2 a_1 B^{1/2} t(x)} |\psi_B|^2\,dx \nonumber\\
&\leq C e^{2B^{1/2} (a_2 D_1^{3/2} - a_1 \epsilon_0)} \| \psi_B \|_2^2
 = {\mathcal O}(B^{-\infty}),
\end{align}
where the last estimate holds if $a_2$ is sufficiently small ($a_2
D_1^{\frac 32} < a_1 \epsilon_0$).

Notice now that the assumption \eqref{eq:genericCurv} implies 
that $\beta \cdot N$ vanishes exactly at order $1$ on $\Gamma$.
Therefore, using the boundedness of $\Omega$,
there exists a constant $C>0$ such that
\begin{align}
C^{-1} d_{\Gamma}(x) \leq \theta(x) \leq C d_{\Gamma}(x),
\end{align}
for all $x \in B(\partial\Omega,2\epsilon_0)$.
Therefore, we observe  that
\begin{itemize}
\item  $\sigma$ is monotone, \item
 $\alpha_1>0$ (from
\eqref{eq:SmallTheta}),
\item  $D$ can be chosen arbitrarily small, 
\item 
 $\lambda_1(B)$ is bounded from above (see \eqref{eq:asymp}),
\end{itemize} 
\noindent and we find
that, if $a_2$ is sufficiently small, there exists $B_0$ such that
\begin{align}
\label{eq:ControlPos}
W_B(x) &- \lambda_1(B) -  \frac{9}{2} a_2^2 B d_{\Gamma}(x) \geq B^{3/4},
\end{align}
for all $x \in B(\partial\Omega,2\epsilon_0)$ with $d_{\Gamma}(x) \geq
D^{-1} B^{-1/4}$, and $B\geq B_0$.

Inserting \eqref{eq:EstNormal} and \eqref{eq:ControlPos} in
\eqref{eq:AlmostDone}, yields, for some constants $C>0$ and $B_0$,
\begin{align}
\label{eq:Final}
\int \chi_1^2(\frac{t}{\epsilon_0}) \chi_2^2(D B^{1/4} d_{\Gamma})
e^{2a_2 B^{1/2} d_{\Gamma}^{3/2}} |\psi_B|^2\,dx
\leq  C e^{ C B^\frac 18}\,,\, \forall B\geq B_0.
\end{align}
Since $e^{2a_2 B^{1/2} d_{\Gamma}^{3/2}}\leq e^{a_2
  2^{5/2}D^{-3/2}\, B^{\frac 18}}$  when $B^{1/4}
d_{\Gamma} \leq \frac 2D$, \eqref{eq:TangentAgmon} follows from \eqref{eq:Final}.
\end{proof}

Consider now the set ${\mathcal M}_{\Gamma}  \subset \Gamma$ where the function $\widetilde{\gamma}_0$ is minimized,
\begin{align}
{\mathcal M}_{\Gamma} := \{ x \in \Gamma \, : \, \widetilde{\gamma}_0 = \widehat{\gamma}_0\}.
\end{align}

\begin{thm}\label{thm:Awaygammahat}~\\
Let $\delta >0$ and suppose that $\Omega \subset {\mathbb R}^3$ satisfies
 Assumptions~\ref{ass:Generic1}, ~\ref{ass:Generic2a} and
 ~\ref{ass:Generic2b}.
 Then, for all $N>0$,  there exists $C_N>0$, such that if $\psi_B$ is a normalized ground state eigenfunction of ${\mathcal H}(B)$, then
\begin{align}
\int_{\{ x \in \Omega \,:\, \dist(x, {\mathcal M}_{\Gamma}) \geq \delta \}} |\psi_B(x)|^2\,dx \leq
C_N B^{-N},
\end{align}
for all $B>0$.
\end{thm}
The proof of the theorem is based on a careful reading of what is
proved, in \cite[Section 15]{HeMo3}\footnote{with the correspondence
 $h=B^{-1}$, $\tau(h)=h^{\delta}$.} in the second zone $\Gamma^{11}$  which corresponds
to the neighborhood of $\Gamma$ defined by
 $$
\Gamma^{11}:=\{x\in \Omega \;|\; 
d_\Gamma (x)+ d(x,\partial \Omega)\leq
 B^{-\delta}\}\,
$$
 where\footnote{Note that $\frac 14 <  \frac{5}{18}$.}  $\delta\in
 ]\frac{5}{18},\frac 13[$. \\
In this region it is shown that we have, for some $\eta>0$,  the inequality
\begin{multline}
B^{\frac 23}\int_{\Gamma^{11}} (\widetilde \gamma_0(x)-\widehat
\gamma_0)|\psi_B(x)|^2\, dx  \\
\leq 
\langle \left(\mathcal H (B)-\lambda(B)\right)
\psi_B\;|\;\psi_B\rangle + C \, B^{\frac 23 -
\eta } \|\psi_B\|^2 \;.
\end{multline}
Then we can proceed like in the previous proof. The only point here
is that the situation is easier because we  need only to have an
estimate in the  region of $\Gamma^{11}$ where $ \dist(x, {\mathcal M}_{\Gamma}) \geq
\delta >0$ (with $\delta$ independent of $B$).

It is clear that, outside this zone, we have already shown that
$\psi_B$ is exponentially small.

\begin{prop}\label{prop:Gauge}~\\
Let $d_{\Gamma}$ be the function defined in Theorem~\ref{thm:TangentAgmon} and let $\Gamma_j$ be one of the curves making up $\Gamma$.
Let $s_0 \in \Gamma_j$ and define, for $\epsilon>0$,
\begin{align}
\Omega(\epsilon, s_0) = \{ x \in \Omega \,:\, \dist(x, \Gamma) < \epsilon \text{ and } \dist(x,s_0)>\epsilon \}.
\end{align}
Then, if $\epsilon$ is sufficiently small, there exists a function
$\phi \in C^{\infty}(\overline \Omega)$ such that the magnetic potential $\widehat{\bf A} := {\bf F}+ \nabla \phi$, satisfies
$$
| \widehat{\bf A}(x)| \leq C\Big(t(x) + d_{\Gamma}(x)^2\Big),
$$
for all $x \in \Omega(\epsilon, s_0)$.
\end{prop}

An easy localization argument shows that we can carry out the above gauge change simultaneously at each $\Gamma_j$.

\begin{cor}\label{cor:GaugeChange}~\\
Let $(s_1,\ldots,s_N) \in \Gamma_1 \times \cdots \times \Gamma_N$ and define, for $\epsilon>0$,
\begin{align}
\label{eq:OmegaEpsilon}
\Omega\big(\epsilon, (s_1,\ldots,s_N)\big) = \{ x \in \Omega \,:\, \dist(x, \Gamma) < \epsilon \text{ and } \min_{j}\dist(x,s_j)>\epsilon \}.
\end{align}
Then, if $\epsilon$ is sufficiently small, there exists a function $\phi \in C^{\infty}(\overline \Omega)$ such that $\widehat{\bf A} := {\bf F}+ \nabla \phi$, satisfies
$$
| \widehat{\bf A}(x)| \leq C\Big(t(x) + d_{\Gamma}(x)^2\Big),
$$
for all $x \in\Omega\big(\epsilon, (s_1,\ldots,s_N)\big)$.
\end{cor}

$\,$
\begin{proof}[Proof of Prop.~\ref{prop:Gauge}]~\\
We use the adapted coordinates $(r,s,t)$ near $\Gamma_j$ defined in \cite[Chapter 8]{HeMo3}. We briefly recall their properties.
Let $\Gamma_j$ be parametrized by arc-length as 
$$
\frac{|\Gamma_j|}{2\pi} {\mathbb S}^1 \ni s \mapsto \Gamma_j(s) \in \partial \Omega.
$$
Given a point $x \in \Omega$ sufficiently close to $\Gamma_j$ there exists a unique point $y(x) \in \partial \Omega$ such that $\dist(x, \partial \Omega) = \dist(x, y(x))$ and a unique point $\Gamma_j(s(x)) \in \Gamma_j$ such that $d_{\Gamma}(y(x)) = \dist_{\partial \Omega}(y(x), \Gamma_j) = \dist_{\partial \Omega}(y(x), \Gamma_j(s(x)))$.
The coordinates $(r,s,t)$ associated to the point $x$ now satisfy
$$
|r| = d_{\Gamma}(y(x)), \quad\quad s=s(x),\quad\quad
t = \dist(x, \partial \Omega) .
$$

Let $\widetilde{A}_1 dr + \widetilde{A}_2 ds+ \widetilde{A}_3 dt$ be the magnetic one-form $\omega_{\bf A} = {\bf A} \cdot d{\bf x}$ pulled-back (or pushed forward) to the new coordinates $(r,s,t)$.
Also write the corresponding magnetic two-form, $d \omega_{\bf A}$, as
$$
\widetilde{B}_{12} dr \wedge ds + \widetilde{B}_{13} dr \wedge dt
+ \widetilde{B}_{23} ds \wedge dt.
$$
Clearly,
\begin{align}
\label{eq:Bij}
\widetilde{B}_{ij} = \partial_i \widetilde{A}_j - \partial_j \widetilde{A}_i,
\end{align}
for $i<j$ and where we identify $(1,2,3)$ with $(r,s,t)$ for the derivatives.

The magnetic field $\beta$ corresponds to the magnetic two-form via the Hodge-map. In particular, since $\beta$ is tangent to $\partial \Omega$ at $\Gamma$ we get that
\begin{align}
\widetilde{B}_{12}(0,s,0) = 0.
\end{align}
We now find a particular solution $\widetilde {\bf A}$ to the equations \eqref{eq:Bij}.
We make the {\it Ansatz}
\begin{align}
\widetilde{A}_1 &= - \int_0^t \widetilde{B}_{13}(r,s,\tau)\,d\tau,\\
\widetilde{A}_2 &= - \int_0^t \widetilde{B}_{23}(r,s,\tau)\,d\tau+ \psi_2(r,s),\\
\widetilde{A}_3 &=0.
\end{align}
Using the relation $d (d\omega_{\bf A})=0$, we see that the above {\it Ansatz} gives a solution if $\psi_2$ is chosen as
\begin{align}
\psi_2(r,s) = \int_0^r \widetilde{B}_{12}(\rho,s,0)\,d\rho.
\end{align}
One verifies by inspection that with these choices
\begin{align}
| \widetilde {\bf A}| \leq C (r^2 + t).
\end{align}
By transporting this $\widetilde {\bf A}$ back to the original coordinates we get the existence of an $\widehat{\bf A}$ with
$$
\curl \widehat {\bf A} = 1,\quad\quad | \widehat{\bf A}(x)| \leq C(t(x) + d_{\Gamma}(x)^2).
$$
Since $\Omega(\epsilon, s_0)$ is simply connected (for sufficiently small $\epsilon$) $\widehat {\bf A}$ is gauge equivalent to ${\bf F}$ and the proposition is proved.
\end{proof}

$\,$
\subsection{The derivative}~\\
In this section we prove how one can derive the monotonicity result from the known asymptotics of the ground state energy and localization estimates for the ground state wave function itself.

Based on these estimates the proof of Theorem~\ref{thm:Derivative} is very similar to the two-dimensional case.

\begin{proof}[Proof of Theorem~\ref{thm:Derivative}]~\\
Let ${\bf F}_{\Omega}$ be the vector potential defined in \eqref{eq:F}. In this proof we will abuse notation slightly and use the symbol ${\mathcal H}(B)$ for the operator $p_{B{\bf F}_{\Omega}}^2$ with Neumann boundary conditions. This corresponds to a gauge change (i.e. a unitary transformation) with respect to the previous notation. In this notation we have
$$
{\mathcal D}({\mathcal H}(B)) = \{ u \in W^{2,2}(\Omega) \,:\, N\cdot \nabla u |_{\partial \Omega} = 0\},
$$
in particular the domain is independent of $B$.
Applying analytic perturbation theory to ${\mathcal H}(B)$ we get the 
existence of $\lambda_{1,\pm}'(B)$. 

Let $\Gamma= \cup_{j=1}^N \Gamma_j$ be the decomposition of $\Gamma$ in disjoint closed curves and let $s_j \in \Gamma_j$ be a point with $\widetilde \gamma(s_j) > \widehat{\gamma}_0$.
Let $\Omega\big(\epsilon, (s_1,\ldots, s_N)\big)$ be as defined in \eqref{eq:OmegaEpsilon} with $\epsilon$ so small that
$$
\widetilde \gamma(x) > \widehat{\gamma}_0
$$
for all $x \in \Omega\big(\epsilon, (s_1,\ldots, s_N)\big)$.
Let $\widehat {\bf A} $ be the vector potential defined in Corollary~\ref{cor:GaugeChange}.

Let  $\widehat Q_B$ the quadratic form
$$
W^{1,2}(\Omega) \ni u\mapsto
\widehat Q_B(u)=  \int_\Omega|(-i\nabla u + B \widehat {\bf A})u|^2 dx\,,
$$
and
$\widehat {\mathcal H}(B)$ be the associated operator. 
Then $\widehat {\mathcal H}(B)$ and ${\mathcal H}(B)$ are unitarily
equivalent: $\widehat {\mathcal H}(B)=e^{iB\phi}{\mathcal H}(B)e^{-iB\phi}$, for some $\phi$ independent of $B$.
With $\psi^+_1(\,\cdot\,;\beta)$ being a suitable choice of normalized
ground state eigenfunction,
we get (by analytic perturbation theory applied to ${\mathcal H}(B)$
and the explicit relation between $\widehat {\mathcal H}(B)$ and ${\mathcal H}(B)$),
\begin{equation}
\lambda_{1,+}'(B) = \langle \widehat {\bf A} \psi^+_1(\,\cdot\,;B) \,,\,  p_{B 
\widehat {\bf A}}
\psi^+_1(\,\cdot\,;B)\rangle +  \langle  p_{B 
\widehat {\bf A}}  \psi^+_1(\,\cdot\,;B) \,,\, \widehat {\bf A}
\psi^+_1(\,\cdot\,;B)\rangle \;.
\end{equation}
We now obtain for any $b >0$, 
\begin{align}
\label{eq:DiffKvot1}
\lambda_{1,+}'(B) & =  \frac{\widehat{Q}_{B+b}(\psi^+_1(\,\cdot\,;B))
  - \widehat{Q}_B (\psi^+_1(\,\cdot\,;B))}{b} 
 - b \int_{\Omega} \vert \widehat {\bf A} \vert ^2 \, |\psi^+_1(x;B)|^2\,dx \nonumber\\
&\geq \frac{\lambda_1(B+b) - \lambda_1(B)}{b} - 
b \int_{\Omega} \vert \widehat {\bf A} \vert ^2 \, |\psi^+_1(x;B)|^2\,dx\,.
\end{align}
We choose $b := B^{\frac{2}{3} - \frac {\eta_0}{ 2}}$, with $\eta_0 <\eta$
 ($\eta$ from
 \eqref{eq:asymp}).
 Then, using \eqref{eq:asymp}, \eqref{eq:DiffKvot1} becomes
\begin{multline}
\label{eq:DiffKvot2}
\lambda_{1,+}'(B)  = \Theta_0 + \widehat{\gamma}_0 B^{-1/3}
\frac{(1+b/B)^{2/3}-1}{b/B} - C B^{-\frac \eta 2} \\
 - b \int_{\Omega} \vert \widehat {\bf A} \vert ^2 \, |\psi^+_1(x;B)|^2\,dx\,,
\end{multline}
for some constant $C$ independent of $ B\geq B_0$.
If we can prove the existence of $C$ such that 
\begin{align}
\label{eq:DecayGauge}
B^{\frac{2}{3}}  \int_{\Omega} \vert \widehat {\bf A} \vert
^2 \, |\psi^+_1(x;B)|^2\,dx \leq C\,,
\end{align}
 then we can take the limit $B\rightarrow \infty$ in \eqref{eq:DiffKvot2} and obtain
\begin{align}
\liminf_{B\rightarrow \infty} \lambda_{1,+}'(B) \geq \Theta_0.
\end{align}

Applying the same argument to the derivative from the left, $\lambda_{1,-}'(B)$, we get (the inequality gets turned since $b<0$)
\begin{align}
\limsup_{B \rightarrow \infty} \lambda_{1,-}'(B) \leq \Theta_0.
\end{align}
Since, by perturbation theory, $\lambda_{1,+}'(B) \leq \lambda_{1,-}'(B)$ for all $B$, we get \eqref{eq:LimitDeriv}.

Thus it remains to prove \eqref{eq:DecayGauge}.
By Corollary~\ref{cor:GaugeChange} we can estimate
\begin{align}
\int_{\Omega} \vert \widehat {\bf A} \vert ^2\; |\psi^+_1(x;B)|^2\,dx
&\leq C \int_{\Omega\big(\epsilon,(s_1,\ldots,s_N)\big)} (t^2+ r^4) |\psi^+_1(x;B)|^2\,dx \nonumber\\
&\quad+
\|\widehat  {\bf A}  \|_{\infty}^2 \int_{\Omega \setminus \Omega\big(\epsilon,(s_1,\ldots,s_N)\big)} |\psi^+_1(x;B)|^2\,dx.
\end{align}
Combining Corollaries~\ref{cor:NormalAgmon} and \ref{cor:AgmonBdry}
 and Theorem~\ref{thm:Awaygammahat},
 we therefore find the existence of a constant $C>0$ such that~:
\begin{align}
\label{eq:Strong}
\int_{\Omega} \vert \widehat {\bf A} \vert ^2\, |\psi^+_1(x;B)|^2\,dx
 \leq C\, B^{-1}\,,
\end{align}
which is stronger than the estimate \eqref{eq:DecayGauge} needed.
\end{proof}

\begin{remark}~\\
Actually, since \eqref{eq:Strong} is stronger than
\eqref{eq:DecayGauge}, 
we do not need the full asymptotics \eqref{eq:asymp} but only the weaker result
$$
\lambda_1(B) = \Theta_0 B + o(B).
$$
Using this information only and choosing $b = \eta B$ instead (and
taking
 the limit $\eta \rightarrow 0$ in the end) we can still achieve
 the proof of Theorem~\ref{thm:Derivative}. However, the second term
 in the 
asymptotics \eqref{eq:asymp} is used in the proof of
Theorem~\ref{thm:Awaygammahat},
 so there seems no gain in not using it at this point.
\end{remark}
\medskip\par
\noindent{\bf Acknowledgements}\\
The two authors were partially supported by the ESF
Scientific Programme in Spectral Theory and Partial Differential
Equations (SPECT).
SF is supported by a Skou-stipend and a Young Elite Researcher Prize from the Danish Research Council. 
Furthermore,  SF wants to thank CIMAT in Guanajuato, Mexico for hospitality during spring 2007.


\begin{thebibliography}{99}

\bibitem[{Ag}]{Ag} S.~Agmon~:
\newblock{\it Lectures on exponential decay of solutions of second order elliptic equations.}
\newblock Math. Notes, T. 29, Princeton University Press (1982).

\bibitem[{Alm}]{Alm} Y.~Almog~:
\newblock  Non-linear surface superconductivity in three dimensions  in the large $\kappa$ limit, 
\newblock Commun. Contemp. Math. 6 (4), p. 637-652 (2004).

\bibitem[{BonFo}]{BonFo} V.~Bonnaillie-No\"{e}l and S.~Fournais~:
\newblock Superconductivity in domains with corners.
\newblock Preprint 2007.

\bibitem[{DaHe}]{DaHe} M. Dauge and  B.~Helffer~:
\newblock Eigenvalues variation I, Neumann problem for Sturm-Liouville
 operators.
\newblock J. Differential Equations  104 (2), p.~243-262 (1993).

\bibitem[{DGP}]{DGP} Q.~Du, M.D. Gunzburger and J.S.~Peterson~:
\newblock Analysis and approximation of the Ginzburg-Landau
 model of superconductivity.
\newblock SIAM Review 34 (1), p. 54-81 (1992).

\bibitem[{DH}]{DutourHelffer} M.~Dutour and B.~Helffer~:
\newblock On bifurcations from normal solutions to superconducting states.
\newblock Rend. Sem. Mat. Univ. Pol. Torino 58, 3 (2000).

\bibitem[{FoHe1}]{FournaisHelffer1} S. Fournais and B.~Helffer~:
\newblock Accurate eigenvalue asymptotics for the magnetic Neumann Laplacian.
\newblock  Ann. Inst. Fourier 56 (1), p.~1-67 (2006).

\bibitem[{FoHe2}]{FournaisHelffer2} S. Fournais and B.~Helffer~:
\newblock On the third critical field in Ginzburg-Landau theory.
\newblock  Comm. Math. Phys. 266 (1) p.~153-196 (2006). 

\bibitem[{FoHe3}]{FournaisHelffer3} S. Fournais and B.~Helffer~:
\newblock Strong diamagnetism for general domains and applications.
\newblock To appear in Ann. Inst. Fourier 2007.

\bibitem[{FoHe4}]{FournaisHelffer4} S. Fournais and B.~Helffer~:
\newblock Optimal Uniform Elliptic Estimates for the Ginzburg-Landau System.
\newblock Preprint 2006.

\bibitem[{GiTr}]{GilbargTrudinger} D.~Gilbarg and N.S.~Trudinger~:
\newblock {\it Elliptic Partial Differential Equations of Second Order.}
\newblock Springer 1998.

\bibitem[{Gi}]{Gi} T.~Giorgi~:
\newblock Superconductors surrounded by normal materials.
\newblock Proc.~Roy.~Soc.~Edingburgh 135 A, p.~331-356 (2005).

\bibitem[{GiPh}]{Giorgi-Phillips} T.~Giorgi and D.~Phillips~:
\newblock The breakdown of superconductivity due to strong fields for the Ginzburg-Landau model.
\newblock SIAM J. Math. Anal. 30 (2), p.~341-359 (1999).

\bibitem[{Hel}]{He2} B.~Helffer~:
\newblock {\it Introduction to the semiclassical analysis 
for the Schr\"odinger operator and applications. }
\newblock Springer Lecture Notes in Math. 1336 (1988).

\bibitem[{He2}]{He3} B.~Helffer~:
\newblock Introduction to semi-classical methods for the Schr\"odinger
operator with magnetic fields.
\newblock CIMPA course in Damas (2004). 
\newblock To appear in the SMF collection  S\'eminaires et Congr\`es  (2007).

\bibitem [{HeMo1}]{HeMo0} B.~Helffer and A.~Mohamed~:
\newblock  Semiclassical analysis for the ground
state energy of a Schr\"odinger operator with magnetic wells.
\newblock J. Funct. Anal. 138 (1), p.~40-81  (1996).

\bibitem [{HeMo2}]{HeMo1} B.~Helffer and A.~Morame~:
\newblock Magnetic bottles in connection with
 superconductivity.
\newblock J. Funct. Anal. 185 (2), p.~604-680 (2001). 

\bibitem [{HeMo3}]{HeMo2} B.~Helffer and A.~Morame~:
\newblock  Magnetic bottles for the Neumann problem: the case of
dimension $3$.
\newblock Proc. Indian Acad. Sci. (Math. Sci.) 112 (1) (2002),
p.~71-84.

\bibitem [{HeMo4}]{HeMo3a} B.~Helffer and A.~Morame~:
\newblock Magnetic bottles for the Neumann problem~: curvature effect
in the case of dimension 3.
\newblock Preprint mp$\_$arc 01-362 (2001).

\bibitem [{HeMo5}]{HeMo3} B.~Helffer and A.~Morame~:
\newblock Magnetic bottles for the Neumann problem~: curvature effect in the case of dimension 3 (General case).
\newblock  Ann. Sci. \'{E}cole Norm. Sup.  37, p.~105-170 (2004).

\bibitem[{HePa}]{He-Pan} B.~Helffer and X-B. Pan~:
\newblock Upper critical field and location of surface nucleation of superconductivity. 
\newblock Ann. Inst. H. Poincar\'e (Section Anal. Non Lin\'eaire) 20 (1), p. 145-181 (2003). 

\bibitem[{LuPa1}]{LuPa1} K.~Lu and  X-B.~Pan~:
\newblock Estimates of the upper critical field for the
Ginzburg-Landau equations of superconductivity.
\newblock Physica D 127, p.~73-104 (1999).

\bibitem[{LuPa2}]{LuPa2} K.~Lu and  X-B.~Pan~:
\newblock Eigenvalue problems of Ginzburg-Landau operator in bounded
domains.
\newblock J. Math. Phys. 40 (6), p.~2647-2670, June 1999.

\bibitem[{LuPa3}]{LuPa3} K.~Lu and  X-B.~Pan~:
\newblock Gauge invariant eigenvalue problems on ${\mathbb R}^2$ and ${\mathbb R}^2_+$.
\newblock Trans. Amer. Math. Soc. 352 (3), p.~1247-1276 (2000).

\bibitem[{LuPa4}]{LuPa5} K.~Lu and  X-B.~Pan~:
\newblock Surface nucleation of superconductivity in $3$-dimension.
\newblock J. Differential Equations 168 (2), p.~386-452 (2000).

\bibitem[{Mon}]{Mon} R.~Montgomery~:
\newblock Hearing the zerolocus of a magnetic field.
\newblock Comm. Math. Phys. 168, p.~651-675 (1995).

\bibitem[{Pan1}]{PanJMP} X-B.~Pan~:
\newblock Superconductivity near critical temperature.
\newblock J. Math. Phys. 44 (6), p.~2639-2678 (2003).

\bibitem[{Pan2}]{Pan3D} X-B.~Pan~:
\newblock Surface superconductivity in $3$ dimensions.
\newblock Trans. Amer. Math. Soc. 356 (10), p.~3899-3937 (2004).

\bibitem[{PaKw}]{PanKwek} X-B.~Pan and K.H.~Kwek~:
\newblock Schr\"{o}dinger operators with non-degenerately vanishing mag\-netic fields in bounded domain.
\newblock Trans. Amer. Math. Soc. 354 (10), p.~4201-4227 (2002).

\bibitem[{S-JSaTh}] {SaSj} D.~Saint-James, G.~Sarma, E.J.~Thomas~:
\newblock {\it Type II Superconductivity.}
\newblock Pergamon, Oxford 1969.

\bibitem [{Te}]{Temam} R.~Temam~:
\newblock {\it Navier-Stokes equation. Theory and numerical analysis.}
\newblock Second printing of 3rd (revised) edition. Elsevier Science Publishers B.V. Amsterdam, 1984.

\bibitem[{TiTi}]{TiTi} D.R.~Tilley and J.~Tilley:
\newblock {\it Superfluidity and superconductivity.}
\newblock  3rd edition. Institute of Physics Publishing, Bristol and Philadelphia 1990.

\bibitem [{Ti}]{T} M. Tinkham, 
\newblock {\it  Introduction to Superconductivity.}
\newblock  McGraw-Hill
Inc., New York, 1975.

\end{thebibliography}
\end{document}